\documentclass[12pt]{article}
\usepackage{epsf}
\usepackage{amsmath}
\def\beq{\begin{equation}}
\def\eeq{\end{equation}}

\def\x{X(3872)}

\begin{document}
\begin{titlepage}
\begin{center}
{\Large \bf William I. Fine Theoretical Physics Institute \\
University of Minnesota \\}
\end{center}
\vspace{0.2in}
\begin{flushright}
FTPI-MINN-07/29 \\
UMN-TH-2619/07 \\
September 2007 \\
\end{flushright}
\vspace{0.3in}
\begin{center}
{\Large \bf Pionic transitions from $X(3872)$ to $\chi_{cJ}$
\\}
\vspace{0.2in}
{\bf S. Dubynskiy \\}
School of Physics and Astronomy, University of Minnesota, \\ Minneapolis, MN
55455 \\
and \\
{\bf M.B. Voloshin  \\ }
William I. Fine Theoretical Physics Institute, University of
Minnesota,\\ Minneapolis, MN 55455 \\
and \\
Institute of Theoretical and Experimental Physics, Moscow, 117218
\\[0.2in]
\end{center}

\begin{abstract}

We consider transitions from the resonance $X(3872)$ to the $\chi_{cJ}$
states of charmonium with emission of one or two pions as a means of
studying the structure of the $X$ resonance. We find that the relative rates for
these transitions to the final states with different $J$ significantly depend on
whether the initial state is a pure charmonium state or a four-quark/molecular
state.

\end{abstract}

\end{titlepage}

\section{Introduction}
The narrow resonance $X(3872)$\,\cite{belle1,cdf,d0,babar1} is an extremely
interesting hadronic object: its mass is tantalizingly close to the $D^0 {\bar
D}^{*0}$ threshold, $w = M(D^0 {\bar D}^{*0})-M(X)=0.6 \pm
0.6\,$MeV\,\cite{cleo}, and the isotopic symmetry is unusually strongly violated
in its decays as follows from the co-existence of the decay  $X(3872) \to \pi^+
\pi^- J/\psi$ on one hand and the decays  $X(3872) \to \pi^+ \pi^- \pi^0 J/\psi$
and $X(3872) \to \gamma J/\psi$\,\cite{belle2,belle3} on the other. The
suggested interpretations of this resonance include a dominantly molecular
state\,\cite{cp,ess,nat,mv1} of charmed mesons $(D^0 {\bar D}^{*0} + D^{*0} {\bar
D}^0)$, possibly of the type discussed in the literature long time
ago\,\cite{ov, drgg}, with an admixture of pure charmonium\,\,\cite{mv2,suzuki},
a dominantly $2^3P_1$ state of charmonium\,\cite{cdfn,mc}, and a virtual
threshold state of $D {\bar D}^*$\,\cite{bugg,yuk,hkkn,mvx}.

A quantitative understanding of the internal structure of this peculiar state is
a challenging task for experimental studies. As has been previously discussed in
the literature\cite{mv1,mv3,cdfn} a measurement of the rates and of the photon
spectrum in the decays $X(3872) \to D {\bar D} \gamma$ can provide a useful
insight into certain features of the structure of $X$.  In particular, it has
been suggested\,\cite{cdfn,mc} that the observed properties of $X(3872)$ can be,
to an extent, mimicked by a $2^3P_1$ state of charmonium, so that any
`molecular' admixture would be viewed as a secondary effect due to the coupling
to the $D {\bar D}^*$ states. In this picture the main available indicator of a
significant isospin violation in $X(3872)$, the approximately equal rate of the
decays $X(3872) \to \rho J/\psi \to \pi^+ \pi^- J/\psi$ and $X(3872) \to \omega
J/\psi \to \pi^+ \pi^- \pi^0 J/\psi$ is explained by the kinematical suppression
of the isospin-allowed transition $X(3872) \to \omega J/\psi$.

In this paper we discuss the transitions from $X(3872)$ to the $\chi_{cJ}$
charmonium states with emission of one or two pions, which can be studied in
addition to the observed processes $X(3872) \to \pi^+ \pi^- J/\psi$ and $X(3872)
\to \pi^+ \pi^- \pi^0 J/\psi$, and which may be instrumental in further
exploration of the $X$ resonance. The discussed  transitions  may be accessible
for experimental observation and may hold the clue to understanding the isotopic
structure of the $X(3872)$ and of the prominence of the four-quark component in
its internal dynamics. We argue that the characteristics of such transitions are
generally completely different between the possible charmonium and molecular
components of the resonance $X(3872)$. The rate of the one-pion transition
relative to the process with two pions is sensitive to the $I=1$ four-quark
component of the $X(3872)$, while the isoscalar four-quark component should give
rise to relative rates of two-pion transitions to the $\chi_{cJ}$ states with
different $J$, which are very likely at variance from those expected for
transitions between charmonium levels.

In Sect.~2 we consider the one- and two-pion transitions for the case where
$X(3872)$ is treated as a $2^3P_1$ state of charmonium. In this consideration we
use the standard approach based on the multipole expansion in
QCD\,\cite{gottfried} and the chiral properties of soft pions\footnote{A recent
discussion of this method in some detail can be found in Ref.~\cite{mv4}.}. We
find that of the isospin-conserving two-pion transitions by far the strongest
should be the decay $\x \to \pi \pi \chi_{c1}$ with the transition to
$\chi_{c2}$ being heavily suppressed and the transition to $\chi_{c0}$ being
forbidden at all in the leading order of the chiral expansion. Given that, the
rate of the strongest transition in absolute terms is expected to be quite small
$\Gamma(\x \to \pi \pi \chi_{c1}) \sim 1\,$keV. The isospin-breaking transitions
$\x \to \pi^0 \chi_{cJ}$ proceeding due to the isotopic symmetry breaking by the
$u$ and $d$ quark masses are estimated to be still weaker than the two-pion
processes. Additionally, the amplitude of the transition $\x \to \pi^0
\chi_{c0}$ turns out to be zero within the considered approach.

A significantly different set of one- and two-pion decay rates is expected in
the case where $X(3872)$ has a substantial four-quark component, so that the
emission of the pions can be considered as a `shake off' of light degrees of
freedom, as discussed in Sect.~3. In this case no strong suppression of a single
pion transition should be expected, and in fact such transitions, proceeding due
to the $I=1$ part of the wave function of $\x$,  are likely to be dominant over
the kinematically suppressed emission of two pions. Moreover, all three
$\chi_{cJ}$ resonances are allowed in the final state, including the
$\chi_{c0}$, which is additionally favored by the phase space factor $p_\pi^3$.

At present we cannot reliably predict the transition rates from a four-quark
state. However a very approximate estimate\,\cite{mv2}, based on the general
understanding of the hadronic parameters, indicates that the  single pion
transitions $\x \to \pi^0 \chi_{cJ}$ should not be very strongly suppressed
relative to the observed decay $\x \to \pi^+ \pi^- J/\pi$. Thus one may hope
that if $\x$ indeed has a significant $I=1$ four-quark/molecular component, the
discussed single-pion transitions may be accessible for experimental studies.

\section{Charmonium}
In this section we treat $X(3872)$ as a charmonium $2 ^3 P_1$ state, so that the
hadronic transitions from this resonance can be studied using the standard
approach based on the multipole expansion. The two-pion transitions arise
\cite{gottfried} in the second order in the $E1$ interaction with the
chromoelectric gluon field $\vec{E}^a$ described by the Hamiltonian

\begin{equation}
H_{E1}=-{1\over 2}\,\xi^a r_i E^a_i(0),
\label{E1}
\end{equation}
where $\xi^a=t_1^a-t_2^a$ is the difference of the color generators acting on
the quark and antiquark, and $\vec{r}$ is the vector of the relative position of
the quark and antiquark. The convention used throughout this paper is that the
QCD coupling $g$ is included in normalization of the gluon field operators.

The one pion transitions $^3 P_1 \to ^3\!\! P_J\, \pi$ are induced by the
interference of the $E1$ interaction~(\ref{E1}) and $M2$ term containing the
chromomagnetic field $\vec{B}^a$ and described by the Hamiltonian \cite{vz}
\begin{equation}
H_{M2}=-{1\over 4 m_Q}\,\xi^a S_k r_j \left(D_j B_k(0)\right)^a,
\label{M2}
\end{equation}
where $D$ is the QCD covariant derivative, $m_Q$ is the heavy quark mass, and
$\vec{S}=\left(\vec{\sigma}_1+\vec{\sigma}_2\right)/2$ is the operator of the
total spin of the quark-antiquark pair.

Generally, one also needs to consider the contribution to the one pion
transition  coming from the Hamiltonian of the form:
\begin{equation}
H_{M2}^{(L)}=-{\xi^a \over 48\, m_Q}\left[L_k\, r_j \, \left(D_j B_k\right)^a
+ \left(D_j B_k\right)^a \, r_j\,L_k\right]\,, \label{HL}
\end{equation}
where $\vec{L}$ is the operator of the orbital momentum of the quark-antiquark
pair. However we will argue later that this contribution cancels due to the
specific form of the gluonic matrix element.

Using the expressions (\ref{E1}) and (\ref{M2}) one can find the one- and two-
pion transition amplitudes in the form
\begin{equation}
A_{\pi\pi}^{(J)} \equiv A \left(\,^3 P_1\to \pi\pi\, ^3 P_J
\right) = \left< \pi \pi \left|\, E_i^a E_j^a \right| 0\right>A_{i
j}^{(J)} \,, \label{ChApi}
\end{equation}
\begin{equation}
A_{\pi}^{\left(J\right)} \equiv  A \left(\,^3 P_1\to \pi\, ^3 P_J
\right)= m_Q^{-1}\left<\pi\left|\,E^a_i\left(D_j
B_k\right)^a+\left(D_j B_k\right)^a E^a_i \right|0\right> A_{i j
k}^{\left(J\right)}\,, \label{ChApipi}
\end{equation}
where the heavy quarkonium amplitudes $A_{i j}^{(J)}$ and $A_{i j k}^{(J)}$ are
defined as follows
\begin{equation}
A_{i j}^{(J)}={1\over 32}{}\left< 1\, ^3\!P_J\left|\,\xi^a r_i\,
\mathcal{G}\, r_j\, \xi^a\right| 2 \, ^3\!P_1 \right>
\label{Aij}
\end{equation}
and
\begin{equation}
A_{i j k}^{(J)}={1\over 128}\left< 1\, ^3\!P_J \left|\, \xi^a\,
r_i \,\mathcal{G}\, r_j\, \xi^a S_k +  r_j\, \xi^a\,\mathcal{G}\,
\xi^a\, r_i\, S_k\right| 2 \, ^3\!P_1\right>
\label{Aijk}
\end{equation}
with $\mathcal{G}$ being the Green's function of the heavy quark pair in a color
octet state. Also in the expressions (\ref{Aij}) and (\ref{Aijk}) we have used
the fact that in the matrix elements between color singlet states the color
factor $\xi^a \ldots \xi^b$ can be replaced by $\left(\delta^{a b}/8\right)
\xi^c \ldots \xi^c$.

In the leading nonrelativistic limit the spin of the heavy quark pair decouples
from the coordinate degrees of freedom. Thus, it is convenient to write the
amplitude in a form with explicitly factorized spin and orbital components. For
this purpose we denote $\zeta_i$ and $\eta_i$ the spin polarization amplitudes
of the initial $^3\!P_1$ and the final $^3\!P_J$ states.  We also introduce the
polarization amplitudes $\phi_i$, $\psi_i$ and $\psi_{ij}$ of the initial state
and the final state with $J=1$ and $J=2$ correspondingly. The tensor $\psi_{i
j}$ is symmetric and traceless, as appropriate for the $J=2$ state. All these
amplitudes are assumed to be normalized in a standard way, so the sums over the
polarization states are defined as
\begin{gather}
\sum_{pol}\zeta_i^*\zeta_j=
\sum_{pol}\eta_i^*\eta_j=\sum_{pol}\phi_i^*\phi_j=
\sum_{pol}\psi_i^*\psi_j=\delta_{ij} \,, \nonumber \\
%\sum_{pol}\phi_{ij}^*\,\phi_{k l}=
\sum_{pol}\psi_{ij}^*\,\psi_{k l}={1\over 2}\left(\delta_{i
k}\,\delta_{j l}+\delta_{i l}\,\delta_{j k}-{2\over 3}\,\delta_{i
j}\,\delta_{k l}\right). \label{polariz}
\end{gather}
Further we define the initial and the final states
\begin{equation}
%\left.\left.\right|\,^3 P_1\, \right>
^3 P_1= \epsilon_{i j\, k}\,\zeta_i\,a_j\,\phi_k/\sqrt{6}\
\,,\qquad
%\left.\left<\,^3 P_J\,\right|\right.
^3 P_J = P_{i j}^{(J)}\,\eta_i\,b_j\,, \label{Chif}
\end{equation}
where $\vec{a}$ and $\vec{b}$ stand for the $P$-wave coordinate parts of the
wave functions and the projection matrices $P_{i j}^{(J)}$ for the definite
total spin $J$ are:
\begin{equation}
P_{i j}^{(0)}=\delta_{i j}/\sqrt{3}\,,\qquad P_{i
j}^{(1)}=\,\epsilon_{i j\, k}\,\psi_k\,/\sqrt{2}\,,\qquad P_{i
j}^{(2)}=\,\psi_{i j}\,. \label{ChP}
\end{equation}

Proceeding to the explicit calculation of the heavy quarkonium matrix elements
we assume as in the heavy quark limit that the spin of the heavy quark pair is
conserved, i.e. $\left<\,\eta_i\,\zeta_j\, \right>=\delta_{i j}$. Generally the
intermediate states contributing to the Green's function in equations
(\ref{Aij}) and (\ref{Aijk}) could be those corresponding to the $S$- and $D$-
wave (the $P$-wave is forbidden by the parity), so that two different structures
are allowed: $\left<b_m \left|\,\xi^a\,r_i\,\mathcal{G}\,
r_j\,\xi^a\,\right|a_n\right>=S\,\delta_{i m}\delta_{j n}+D\left(\delta_{i
j}\,\delta_{m n}+\delta_{i n}\,\delta_{j m}-2/3\,\delta_{i m}\,\delta_{j
n}\right)$. However, assuming that the relevant intermediate states are
separated by a large energy gap $\Delta$ from the discussed $P$-wave states, one
can approximate the Green's function $\mathcal{G}$ as being proportional to the
unit operator: $\mathcal{G}\simeq 1/\Delta$. In this case the coefficient in
front of the $S$ wave is related to that of the $D$ wave: $S=5/3 D$. Thus within
the introduced notations the heavy quarkonium amplitudes can be written in
terms of a scalar quantity $\mathcal{A}$
\begin{equation}
A_{i j}^{(J)}= \mathcal{A} \left( \delta_{i n}\delta_{j
p}+\delta_{i p}\delta_{j n}+\delta_{i j}\delta_{n
p}\right)\epsilon_{m p r}\,P_{m n}^{(J)}\, \phi_r\,, \label{ChAij}
\end{equation}
\begin{equation}
A_{i j k}^{(J)}= {i \mathcal{A} \over 2}\left( \delta_{i
n}\delta_{j p}+\delta_{i p}\delta_{j n}+\delta_{i j}\delta_{n
p}\right)\epsilon_{q m k}\epsilon_{q p r}\,P_{m n}^{(J)}\,
\phi_r\,.
\label{ChAijk}
\end{equation}
The quantity $\mathcal{A}$ depends on details of the heavy quarkonium dynamics
and at present is highly model dependent, however, it is clear that this
quantity cancels in the considered ratios of the rates of the single pion and
two-pion transitions.

The gluonic matrix element for the one pion transition in the Eq.(\ref{ChApi})
can be written using the form described in \cite{mv5}
\begin{equation}
i\left<\pi\left|\,E^a_i\left(D_j B_k\right)^a+\left(D_j
B_k\right)^a E^a_i \right|0\right>= {X\over 15}
\left(3p_j\delta_{ik}-p_i\delta_{jk}\right),
\label{EDB}
\end{equation}
with the form factor $X$ related to the well known expression
\cite{gtwn}
\begin{equation}
X=\left<\pi\left|\,G^a \tilde{G}^a \right|0\right>=8 \pi^2
\sqrt{2} \,\,{m_d-m_u \over m_d+m_u}\, f_\pi m_\pi^2 \,, \label{X}
\end{equation}
where $m_u$ and $m_d$ are the masses of the $u$ and $d$ quarks and $f_\pi\approx
130\,$MeV. It should be noted that the specific form of the gluonic matrix
element Eq.(\ref{EDB}), namely the absence of the term proportional to
$\delta_{i j}$, makes the contribution of the orbital chromomagnetic term
(\ref{HL}) being proportional to the structure $(\vec{L} \cdot \vec{r})$ which
is obviously equal to zero.

Using the expressions (\ref{EDB}), (\ref{X}) and (\ref{Chif})-(\ref{ChAijk}) one
readily finds the squares of the transition amplitudes summed over the
polarizations of the final $^3 P_J$ state:
\begin{equation}
\overline{|\, A_\pi^{(2)}|\,^2}= {X^2\, p_\pi^2\over 15\,m_Q^2}|
\,\mathcal{A}|^2\,, \qquad \overline{|\, A_\pi^{(1)}|\,^2}={X^2\,
p_\pi^2\over 9\,m_Q^2}|\,\mathcal{A}|^2\,,\qquad \overline{|\,
A_\pi^{(0)}|\,^2}=0\,, \label{ChApi2}
\end{equation}
where $p_\pi=|\,\vec{p}\,|$ is a pion momentum. We note here that the vanishing
of the amplitude $A_\pi^{(0)}$ is a result of our approximation for the Green's
function as being proportional to the unit operator. The ratio of the amplitudes
(\ref{ChApi2}) then gives the ratio of the decay rates:
\begin{equation}
\Gamma_2\,:\Gamma_1\,:\Gamma_0= 3 p_{\pi\,(2)}^{\,3}\,: 5
p_{\pi\,(1)}^{\,3}\,:0\approx 1:\,2.70 :0\,,
\end{equation}
where $p_{\pi\,(1)}\approx 334$ MeV and $p_{\pi\,(2)}\approx 285$ MeV are the
pion momenta in the transitions to the $^3 P_1$ and $^3 P_2$ states
correspondingly.

Let us proceed now to evaluating the two pion transition rates. Using general
arguments based on the isospin and parity considerations one can see that the
pion pair in the transitions $^3 P_1 \to\, ^3 P_J\, \pi\pi$ could only be
produced in the even partial waves. From the properties of the chiral algebra it
follows that the expansion of the $S$- and $D$-wave amplitudes starts from the
terms quadratic in the pion momentum (and mass) and therefore kinematically
these amplitudes are both of the same order in the soft pion limit.

Considering the transition $1^{++}\to 0^{++}$ one can easily see that the pion
pair besides the $D$-wave motion in its center of mass frame should also be
involved in a $D$-wave motion as a whole relatively to the $0^{++}$ state, thus
the discussed amplitude has to be proportional to the fourth power of the pion
momentum in the chiral expansion. Taking into account the small energy release
in the transitions $2\, ^3\! P_1\to\, 1 ^3 P_J\,\pi\pi$ we should thus expect a
strong suppression of the transition to the $\chi_{c0}$ state in comparison with
the transitions to the $\chi_{c1}$ and $\chi_{c2}$.

The amplitudes of the two pion transitions to the $\chi_{c1}$ and $\chi_{c2}$
can readily be evaluated using the general form~\cite{vz,ns} of the amplitude of
the dipion creation by the chromoelectric field (\ref{ChAij})
\begin{equation} \left<\pi^+\pi^-\left|E_i^a
E_j^a\right|0\right>={8\pi^2\over 3 b}\left(q^2+m^2_\pi\right)
\delta_{ij}+{12 \pi^2\over b}\, \kappa \left[ \, p_{1 i} \,
p_{2 j}+p_{1 j} \, p_{2 i}-\left(\varepsilon_1
\varepsilon_2+\vec{p}_1 \cdot \vec{p}_2\right)\delta_{i j} \, \right]\,,
\label{EE}
\end{equation}
where $\varepsilon_{1,\,2}$ and $\vec{p}_{1,\,2}$ are the energy and momentum of
each pion, $\,q=p_1 + p_2$ is the total 4-momentum of the pion pair; $b=9$ is
the first coefficient in the beta function for QCD with three quark flavors, and
$\kappa$ is a parameter introduced in \cite{ns}. The experimental value of the
parameter $\kappa\approx 0.2$ \cite{bes} agrees with the original theoretical
estimate \cite{ns}.

The first term in the Eq.(\ref{EE}) arising from the conformal anomaly
\cite{mv4} describes the $S$ wave production of the two pions in their c.m.
frame
while the term proportional to $\kappa$ corresponds to both $S$- and $D$-waves.
For an approximate estimate of the amplitude of the transition $2\, ^3 P_1\to 1
^3 P_1 \pi\pi$ we neglect the contribution of the $\kappa$-term since this
process is dominated by the large $S$-wave contribution arising from the first
term in Eq.(\ref{EE}). On the other hand for the transition $2\, ^3 P_1\to 1 ^3
P_2 \pi\pi$ there is no $S$-wave pion pair emission and only the $D$-wave part
of the $\kappa$-term contributes to the amplitude. In this way one estimates
\begin{equation}
\overline{ |\,A_{\pi\pi}^{(1)}|\,^2}= 150 \left({8\pi^2\over 3
b}\right)^2
\left(q^2+m_\pi^2\right)^2\,\left|\,\mathcal{A}\right|\,\!^2\,,
\label{ChA1pipi2}
\end{equation}
\begin{equation}
\overline{ |\, A_{\pi\pi}^{(2)}|\,^2}=12\left({12\pi^2\over
b}\,\kappa \right)^2\left[\vec{p}_1\!^2\, \vec{p}_2\!^2+{1 \over
3}\left(\vec{p}_1\cdot \vec{p}_2 \right)^2
\right]\left|\,\mathcal{A}\right|\,\!^2\to
\frac{40}{3}\left({12\pi^2\over b}\,\kappa \right)^2\,
\vec{p}_1\!^2\, \vec{p}_2\!^2 \left|\,\mathcal{A}\right|\,\!^2,
\label{ChA2pipi2}
\end{equation}
where in the last transition in the Eq.(\ref{ChA2pipi2}) the averaging on the
relative angle between the momenta was made.

For estimates of the rates of $\pi\pi$ transitions one needs to evaluate the
corresponding phase space integrals at the energy release $\Delta^{(J)}=M( 2 ^3
P_1) - M( 1 ^3 P_J)$
\begin{equation}
W_{\pi\pi}^{(J)}=\int\overline{|\, A_{\pi\pi}^{(J)}|\, ^2}\,\,
2\pi\,
\delta\left(\Delta^{(J)}-\varepsilon_1-\varepsilon_2\right){d^3
p_1\over (2\pi)^3 \,2 \varepsilon_1} {d^3 p_2\over (2\pi)^3 \,2
\varepsilon_2}\,.
\label{ChWpipi}
\end{equation}
 A numerical
integration  using the expressions (\ref{ChA1pipi2}) and
(\ref{ChA2pipi2}) then yields
\begin{equation}
{\Gamma \left( 2\, ^3 P_1\to \chi_{c1}\, \pi\pi \right)\over
\Gamma \left( 2\, ^3 P_1\to \chi_{c2}\, \pi\pi\right)}\approx
 10^4~,
\label{rat21}
\end{equation}
so that the transition $2 ^3 P_1\to \chi_{c2}\,\pi\pi$ is quite small in
comparison to the $2 ^3 P_1\to \chi_{c1}\,\pi\pi$ one. Clearly, the main factor
responsible for such strong dominance of the transition to the $\chi_{c1}$ state
is the enhancement by the conformal anomaly of the $S$ wave production of the
pions in the amplitude described by Eq.(\ref{EE}).

Using the expression (\ref{ChApi2}) and the value of the charmed quark mass $m_c
\approx 1.4\,$GeV, one can also find the ratio of one- and two-pion decay rates
\begin{equation}
{\Gamma \left( 2\, ^3 P_1\to \chi_{c1} \pi^0 \right)\over \Gamma
\left( 2\, ^3 P_1\to \chi_{c1} \pi^+ \pi^-\right)}\approx 0.04\,.
\end{equation}
One can see from this estimate that for the case where $X(3872)$ is considered
to be a charmonium $2 ^3 P_1$ state the one-pion transition is significantly
suppressed relative to the two pion one as a result of small isospin violation.

The absolute value of the discussed decay rates can be very approximately
estimated by using the known rate of the decay $\psi\left(2S\right)\to
J/\psi(1S)\,\pi^0$. Within the similar description \cite{is} the gluonic matrix
element of the latter decay is found as
\begin{equation}
{1 \over 64}\left< \eta_m\, \psi_m \left|\,\xi^a\,
r_i\,\mathcal{G}\, r_j\,\xi^a\,S_k \right| \zeta_n\, \phi_n
\right>/3= {5 i \over \sqrt{6}}\,\mathcal{A}_s\, \delta_{i
j}\,\epsilon_{k m n} \,\psi_m^*\,\phi_n\,,
\end{equation}
so that the rate of the decay is given by
\begin{equation}
\Gamma={4 \,X^2\over 27 m_Q^2}{p_\pi^3\over
2\pi}|\,\mathcal{A}_s|^2\,,
\end{equation}
where the quantity $\mathcal{A}_s$ is an analog, relevant for the $2 ^3 S_1
\to 1 ^3 S_1$ transition, of the amplitude $\mathcal{A}$ introduced in
Eqs.(\ref{ChAij}) and (\ref{ChAijk}). Although $\mathcal{A}_s$ and $\mathcal{A}$
arise from overlap integrals between different pairs of charmonium levels, we
consider them as being of the same order, $\mathcal{A}\sim\mathcal{A}_s$, in
lieu of a more reliable understanding. Proceeding in this way, we estimate
\begin{equation}
\Gamma\left( 2 ^3 P_1\to \chi_{c1}\,\pi^0\right) \sim {3 \over 4}
\left({p_{\pi(1)}\over p_\pi}\right)^3\,\Gamma\left(\psi'\to
J/\psi\,\pi^0\right)\simeq 0.06\, \mbox{keV}\,,
\end{equation}
where $p_\pi\approx 574$~MeV is the pion momentum in the $\psi(2S)$ decay. Even
with a large uncertainty in this estimate one can conclude that the rate of the
strongest of the discussed transitions from $X(3872)$ treated as charmonium $2\,
^3 P_1$ state, $ 2 ^3 P_1 \to 1 ^3 P_1\,\pi\pi$, is expected to be very small.

\section{Molecular state}
Another possible scenario is to consider the $X(3872)$ as being a four quark
state made from the two heavy and two light quarks. In this case it appears
reasonable to treat the discussed transitions $X \to \pi \chi_{cJ}$ and $X \to
\pi \pi \chi_{cJ}$ as a `shake off' of the light quarks. In particular, the spin
dependent `heavy-light' quark interaction is proportional to the inverse
power of the heavy quark mass $m_Q^{-1}$, so that any exchange of the
polarization between the light and heavy degrees of freedom is expected to be
suppressed. Neglecting such exchange the total wave function can be written as a
product of the two wave functions describing the heavy and light degrees of
freedom separately. For this purpose we define the polarization amplitudes
$h_i^{(1,\,2)}$ and $l_i^{(1,\,2)}$ associated with the heavy and the light
quark pairs in the initial and the final states. For the states with the
definite $J$ we introduce the polarization amplitudes $\phi_i$ and $\psi_i$
corresponding to the initial and final states with $J=1$ and the symmetrical and
traceless tensor amplitude $\psi_{i j}$ for the final $^3 P_2$ state. All these
amplitudes are assumed to be normalized in the standard way and sums over
polarizations are the same as in Eq.(\ref{polariz}). Finally, introducing the
operators $L_i^{(1,\,2)}$ and $H_i^{(1,\,2)}$  acting on the polarization
amplitudes $l_i$ and $h_j$: $\langle
h_i^{(2)}|\,H_s^{(2)}\,H_s^{(1)}|\,h_j^{(1)}
\rangle \,\sim \delta_{i j}$ and $\langle
l_i^{(2)}|\,L_m^{(2)}\,L_n^{(1)}|\,l_j^{(1)}
\rangle \,\sim\delta_{i m}\delta_{j n}$ one can write the amplitude for the one
pion
transition
\begin{equation}
A_{\pi}^{(J)}=\left<\, ^3 P_J\left| p_{\,l}\, \epsilon_{l m n}\,
L_m^{(2)}\, L_n^{(1)}\,H_s^{(2)}\,H_s^{(1)} \right|\, ^3
P_1\,\right>
\label{XApi}
\end{equation}
with the initial and the final states
\begin{equation}
^3 P_1 = \epsilon_{i j k}\, h_j^{(1)}\, l_k^{(1)}\,\phi_i/\sqrt{6}
\,,\qquad ^3 P_J = P_{i j}^{(J)}\,h_i^{(2)}\, l_j^{(2)}\,,
\label{Xif}
\end{equation}
where the projection matrices $P_{i j}^{(J)}$ are the same as defined in
Eq.(\ref{ChP}). Using these relations one can readily find the amplitudes of the
one pion transition to the states with the different total spin:
\begin{equation}
A_{\pi}^{(0)}= 2 \, p_i\,\phi_i\, \mathcal{A}_1/ \sqrt{3}\,,\qquad
A_{\pi}^{(1)}=\,\epsilon_{i j k}\,\phi_i\,\psi_j^*
\,p_k\,\mathcal{A}_1/ \sqrt{2}\,,\qquad A_{\pi}^{(2)}=-
\,\phi_i\,\psi_{i j}^*\,p_j\,\mathcal{A}_1\,.
\label{XApi1}
\end{equation}
The quantity $\mathcal{A}_1$ accounts for the internal dynamics and at present
can only be discussed on a model dependent basis. Taking the square of these
expressions and summing over the polarization amplitudes of the final states
complete the routine procedure of finding the square of the matrix element
\begin{equation}
\overline{|\,A_{\pi}^{(0)}|\, ^2}={4\over 3}\, p_\pi^2 |\,
\mathcal{A}_1|^2\,,\qquad\overline{|\,A_{\pi}^{(1)}|\, ^2}=
p_\pi^2 |\, \mathcal{A}_1|^2\,,\qquad\overline{|\,A_{\pi}^{(2)}|\,
^2}={5\over 3} \, p_\pi^2 \, |\, \mathcal{A}_1|^2 \,.
\label{XApi2}
\end{equation}
This relation readily gives us the ratio of the partial decay
widths for the one pion transition
\begin{equation}
\Gamma_0\,:\Gamma_1\,:\Gamma_2= 4 p_{\pi\,(0)}^{\,3}\,: 3
p_{\pi\,(1)}^{\,3}\,:5 p_{\pi\,(2)}^{\,3}\approx 2.88:0.97 :1\,,
\label{rat012}
\end{equation}
where $p_{\pi\,(0)}\approx 436\,$MeV is the pion momentum in the
transition to the $\chi_{c0}$ state.

For the two-pion transition from a four-quark state essentially the only
guidance is provided by general chiral properties, which require that the
expansion of the amplitude in the pion 4-momenta starts with a bilinear term. In
this order only $S$ and $D$ waves for the dipion are possible, and one can write
the amplitude in terms of the explicit partial wave decomposition:
\begin{equation}
A_{\pi\pi}^{(J)}=\left<\, {^3 P_J} \left|\, L_m^{(2)}\,
L_n^{(1)}\,\left(\,\delta_{m n} \, S+\Phi_{m n}\, D\right)H_s^{(2)}\,H_s^{(1)}\,
\right| {^3
P_1}\, \right>\,,
\label{XApipi}
\end{equation}
where  the traceless $D$-wave tensor $\Phi_{i j}$ is defined in a standard way:
$\Phi_{i
j}= p_{1\,i}\,p_{2\,j} + p_{1\,j}\,p_{2\,i} - 2/3\,\delta_{i
j}\left(\vec{p}_1\cdot\vec{p}_2\right)$. In the leading order of the chiral
expansion the $S$ wave amplitude can generally be written as $S= a\, q^2 + b \,
\varepsilon_1 \varepsilon_2 + c \, m_\pi^2$ with $a$, $b$ and $c$ being  
coefficients generally of order one, so that there is in fact no
kinematical suppression of the $D$-wave in comparison with the $S$-
wave, and both these terms enter the amplitude on equal footing. It can be also
noted, once again, that no amplitude for transition to the $1 ^3 P_0$ state
arises in this order of expansion.

One finds from the expressions (\ref{Xif}) and (\ref{XApipi}) the amplitudes of
the two pion transitions in the form
\begin{equation}
A_{\chi_{c0}}=0\,,\qquad A_{\chi_{c1}}=\left(2 \delta_{i
j}\,S-\Phi_{i j}\,D\right)\phi_i\,\psi_j^*\,\mathcal{A}_2
/\sqrt{2}\,,\qquad A_{\chi_{c2}}=D \, \epsilon_{i j\,
k}\,\phi_i\,\psi_{j\, l}^*\Phi_{l k}\,\mathcal{A}_2\,,
\end{equation}
where the quantity $\mathcal{A}_2$ describes the internal properties of the
transition and at present is model dependent. Using these expressions and
following the standard prescription it is straightforward to find the averaged
squares of the transition amplitudes
\begin{equation}
 \overline{|\,A_{\pi\pi}^{(1)}|\, ^2}= |\,
\mathcal{A}_2|\,^2 \left\{ 2 \,|\,S
|\,^2+{1 \over 3}\, \left[ \vec{p}_1\!^2\,\vec{p}_2\!^2+{1\over
3}\left(\vec{p}_1\cdot \vec{p}_2
\right)^2\right ] |\,D|\,^2\right\}~,
\label{XA1pipi2}
\end{equation}
\begin{equation}
\overline{|\,A_{\pi\pi}^{(2)}|\, ^2}= |\, \mathcal{A}_2|^2 \,
\left[ \vec{p}_1\!^2\,\vec{p}_2\!^2+{1\over
3}\left( \vec{p}_1\cdot \vec{p}_2 \right)^2\right ] |\,D|\,^2~.
\label{XA2pipi2}
\end{equation}
As previously discussed, in this case there is generally no reason to expect an
enhancement of the $S$ wave over the $D$ wave, so that the ratio of the two
decay rates may be not as dramatic as in Eq.(\ref{rat21}).

\section{Summary}
The described estimates of the rates of the pionic transitions from $\x$ to the
$\chi_{cJ}$ states of charmonium lead us to conclude that these rates exhibit
significantly different patterns depending on the internal structure of the
$\x$. Namely, if this resonance is dominantly a charmonium $^3P_1$ state, the
pionic transitions are quite weak with the process $\x \to \pi \pi \chi_{c1}$
being dominant. The single pion transitions, proceeding through the isospin
breaking by the light quark masses, are even more suppressed with the amplitude
of the decay $\x \to \chi_{c0}$ vanishing altogether as a result of a special
form of the matrix element in Eq.(\ref{EDB}). If however $\x$ is a
four-quark/molecular state with a significant isovector part in its wave
function, the considered single pion transitions should be greatly enhanced and
exhibit comparable rates for all three $\chi_{cJ}$ resonances in the final
state, as described by Eq.(\ref{rat012}). One can notice that the yield of the
$\chi_{c0}$ in the final state in this case is the highest, while in the pure
charmonium case this yield should be strongly suppressed. For the transitions
with emission of two pions in the case of dominantly molecular $X(3872)$ we find
that generally there is no reason to expect a very strong dominance of the
production of the $\chi_{c1}$ resonance in the final state as compared to
$\chi_{c2}$. The production of the $J=0$ state, $\chi_{cJ}$, is expected to be
strongly suppressed in either case due to the general properties of the
soft-pion expansion. The relative rate of the single and double pion emission is
practically impossible to estimate quantitatively at present for a molecular
$\x$. At a qualitative level, once the isospin suppression of the single pion
processes is removed by the four-quark structure of $\x$, one might expect a
larger rate for the one pion transitions than for the two-pion processes, given
the small energy release in the considered decays.

\section*{Acknowledgment}
The work of M.B.V. is supported in part by the DOE grant DE-FG02-94ER40823.

\end{document}